\pdfoutput=1

\documentclass[
 reprint,
superscriptaddress,
 amsmath,amssymb,
 aps,
]{revtex4-2}

\usepackage[table,dvipsnames]{xcolor}
\usepackage{dcolumn}% Align table columns on decimal point
\usepackage{bm}% bold math
\usepackage[T1]{fontenc}
\usepackage{braket}
\usepackage{amsmath}
\usepackage{amssymb}
\usepackage{nccmath}
\usepackage{mathrsfs}
\usepackage{multirow}
\usepackage{siunitx}
\usepackage{graphicx}
\emergencystretch=\maxdimen
\begin{document}
\title{Broadband plasma spray anti-reflection coating technology for millimeter-wave astrophysics}

\author{Oliver Jeong}
\email[]{objeong@berkeley.edu}
\affiliation{Department of Physics, University of California, Berkeley, CA 94720, USA}

\author{Richard Plambeck}
\affiliation{Radio Astronomy Laboratory, University of California, Berkeley, CA 94720, USA}

\author{Christopher Raum}
\affiliation{Department of Physics, University of California, Berkeley, CA 94720, USA}
\affiliation{Radio Astronomy Laboratory, University of California, Berkeley, CA 94720, USA}

\author{Aritoki Suzuki}
\affiliation{Physics Division, Lawrence Berkeley National Laboratory, Berkeley, CA 94720, USA}

\author{Adrian T. Lee}
\affiliation{Department of Physics, University of California, Berkeley, CA 94720, USA}
\affiliation{Radio Astronomy Laboratory, University of California, Berkeley, CA 94720, USA}
\affiliation{Physics Division, Lawrence Berkeley National Laboratory, Berkeley, CA 94720, USA}

\begin{abstract}	
We present a broadband plasma spray anti-reflection (AR) coating technology for millimeter-wave astrophysics experiments with large-format, cryogenic optics. By plasma spraying alumina- and silica-based powders, we have produced coatings of tunable index of refraction and thickness, low loss, and coefficient of thermal expansion matched to alumina substrates. We demonstrate two-layer AR coatings on alumina with reflection below 5\% over 82\% and 69\% fractional bandwidths for 90/150 and 220/280 $\si{\GHz}$ passband designs, respectively, and band-averaged absorption loss reduced to $\sim$1\% at 100 $\si{\kelvin}$ for both AR coatings. We describe the design, tolerance, fabrication process, and optical measurements of these AR coatings.

\vspace{4mm}

\noindent \copyright\ 2023 Optica Publishing Group. One print or electronic copy may be made for personal use only. Systematic reproduction and distribution, duplication of any material in this paper for a fee or for commercial purposes, or modifications of the content of this paper are prohibited.
\end{abstract}

%\date{\copyright\ 2023 Optica Publishing Group. One print or electronic copy may be made for personal use only. Systematic reproduction and distribution, duplication of any material in this paper for a fee or for commercial purposes, or modifications of the content of this paper are prohibited} 

\maketitle

\section{Introduction}
\label{sec:intro}
The rapidly increasing scale of photon noise limited superconducting detector arrays for millimeter-wave astrophysics has necessitated the development of high-throughput optical systems with large, diffraction-limited field of view. Sensitivity can be improved by increasing the physical scale of focal planes and the frequency coverage of each pixel using multi-chroic detectors and broadband optics. Increased frequency coverage also enables spectral characterization of galactic foregrounds for component separation~\cite{planck2018foreground}. Alumina and silicon are common cryogenic lens material for these systems due to their high index of refraction (n $\approx$ 3)~\cite{rajab2008, wollack2020}, low loss-tangent (tan $\delta$ $<$ $10^{-4}$)~\cite{vila1998, parshin1995}, and relatively high thermal conductivity ($k\sim1$ $\si{\watt{\meter}^{-1}{\kelvin}^{-1}}$ for alumina~\cite{nist} and $\sim200$ $\si{\watt{\meter}^{-1}{\kelvin}^{-1}}$ for silicon~\cite{thompson1961} at 4 K). The resulting reflection of $\sim$30$\%$ per vacuum-dielectric interface requires the application of an appropriate anti-reflection (AR) coating.

\par

An AR coating is formed by placing one or more dielectric layers on a vacuum-dielectric boundary of a refractive optical element. Using the principle behind quarter-wave impedance matching, each coating layer's index of refraction and thickness are tuned such that reflected light at every boundary destructively interferes with each other. As many modern cosmic microwave background (CMB) experiments utilize multi-chroic detectors to simultaneously observe multiple frequency bands with a single receiver, it is crucial the optics possess appropriately wide passbands by utilizing multi-layer AR coatings. While increasing the number of layers of a well-optimized, quarter-wavelength AR coating widens the bandwidth, there are trade-offs of greater scattering and absorption loss, as well as practical concerns of increased complexity, cost, and lead time of fabrication~\cite{suzuki}. A good rule of thumb in AR coating design is to determine the minimum number of layers which maximizes the optical efficiency.

\par

Fabrication and cryogenic operation of multi-layer AR coatings present a significant challenge. Each layer has a unique index of refraction and thickness, requiring fine tunability of both parameters for flexible optimization of the transmission passband for many different experiments. The loss tangent must be sufficiently low for every layer to minimize the dissipative loss of the optics. Cryogenic operation of refractive optics requires well matched coefficient of thermal expansion (CTE) between the layers of the dielectric stack to mitigate delamination from differential thermal contraction. Multiple AR coating technologies are available for millimeter-wave applications using plastic layers~\cite{nadolski, hargrave}, metal mesh layers~\cite{pisano1, pisano2}, epoxy-based layers~\cite{rosen}, or sub-wavelength metamaterial structures~\cite{takaku, datta, golec}. However, these existing technologies involve a complicated fabrication process~\cite{pisano1, pisano2, rosen, takaku, datta, golec} or present significant risk of cryogenic delamination~\cite{nadolski, hargrave, rosen}. We present the application of plasma spraying to construct an AR coating solution which demonstrates high optical efficiency for larger aperture, broadband optics with robust adhesion at cryogenic temperatures. Plasma spray is a thermal coating process with common applications in thermal, mechanical, chemical, and electrical barriers for small electronic to large aerospace components. This fabrication process is precise, straightforward, fast, and scalable. 

\par

This paper advances on the work of Suzuki~\cite{suzuki} and Jeong et al.~\cite{jeong} with an end-to-end plasma spray process for multi-layer AR coatings on large aperture optics. We have expanded the range of achievable index of refraction such that a PTFE laminate is no longer necessary for a multi-layer AR coating utilizing plasma spray. Furthermore, we have demonstrated a precise and accurate coating deposition process over large diameter, optical surfaces. This technology has been developed for the Simons Array~\cite{stebor} and SPT-3G experiments~\cite{sobrin} but is suitable for any cryogenic millimeter-wave applications with high index refractive optics. In Section~\ref{sec:req} we detail the design requirements for the Simons Array optics, which are applied to generate the AR coating design described in Section~\ref{sec:design}. In Section~\ref{sec:fab} we describe the fabrication process of these AR coatings on large-aperture, plano-convex alumina lens surfaces and the resulting thickness characterization. Section~\ref{sec:setup} gives an overview of the measurement setup and method for millimeter wavelengths at ambient and cryogenic temperatures. In Section~\ref{sec:performance}, we present transmission measurements of plasma spray AR coated alumina.

\section{Requirements}
\label{sec:req}

The AR coating performance requirements detailed in this work are determined by the optical design goals of the Simons Array experiment which aims to measure the faint, polarized signals of the CMB. Simons Array consists of three cryogenic receivers, each with its own re-imaging optics and kilo-pixel focal plane bolometer array. Two receivers, POLARBEAR-2a and POLARBEAR-2b, contain multi-chroic bolometers designed to observe at 90 (75--104) and 150 (129--167) $\si{\GHz}$ bands simultaneously. The third receiver, POLARBEAR-2c, is designed to observe at 220 (198--242) and 280 (243--297) $\si{\GHz}$ bands simultaneously. Each receiver contains three convex re-imaging lenses and one infrared filter plate made of high purity (99.9\%) alumina for f/1.9 optics to illuminate its large focal plane. These alumina optics are produced by CoorsTek, Inc., each with $\sim$500 $\si{\mm}$ optical diameter and 0.005 $\si{\mm}$ precision to design. The goal for Simons Array AR coating is to suppress total band-averaged reflection and absorption loss of an AR coated optical element to $\sim$1\%, each. Finally, the coating must be robust against cryogenic thermal cycling. CMB experiments operate with refractive optics at temperatures below 50 $\si{\kelvin}$ to reduce absorptive loss and thermal emission. Coatings without sufficiently matched CTE to alumina and silicon lenses will delaminate upon thermal cycling. This becomes more difficult with multi-layer coatings for broadband optics with more dielectric layers to match in CTE.   

\begin{figure}[htbp]
\centering
\includegraphics[width=0.98\linewidth]{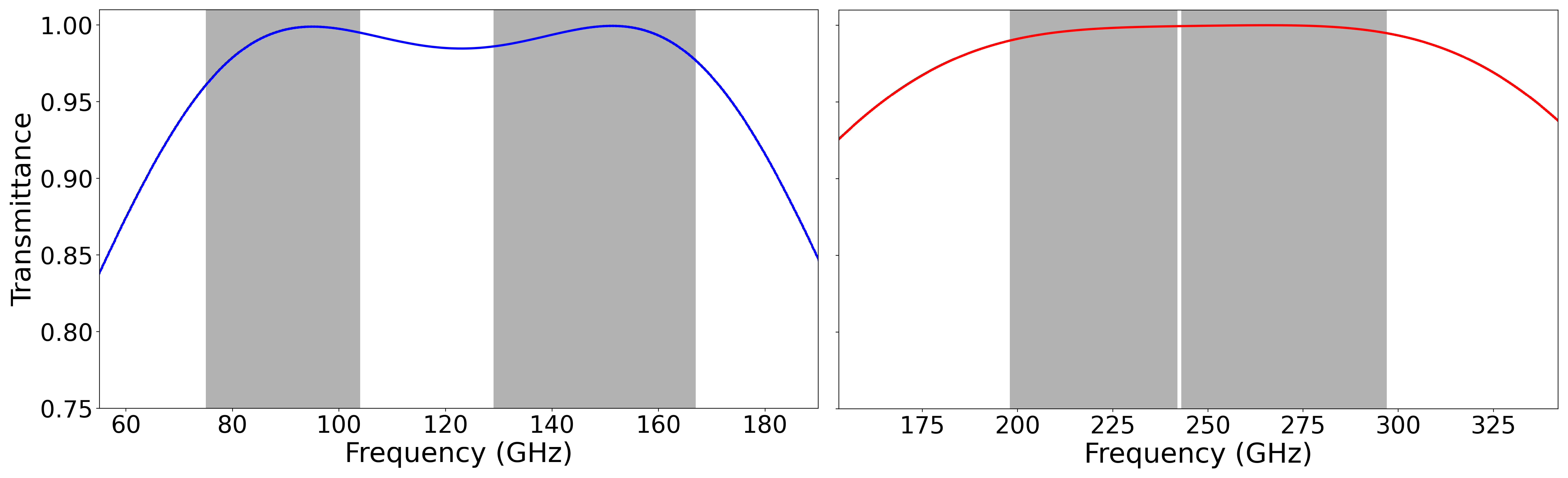}
\caption{Theoretical transmittance spectra of alumina AR coatings for optimal performance at 90/150 $\si{\GHz}$ bands (left) and 220/280 $\si{\GHz}$ bands (right).}
\label{fig:opt_theory}
\end{figure}

\begin{table}[htbp]
    \centering
    \begin{tabular}{|c|c|c|c|c|c|}
    \hline
    \multirow{2}{*}{\textbf{Passband}} & \multicolumn{2}{c|}{\textbf{Top layer}} & \multicolumn{2}{c|}{\textbf{Bottom layer}} & \multirow{2}{*}{\textbf{R (\%)}} \\
    \cline{2-5}    
     & n & t ($\si{\um}$) & n & t ($\si{\um}$) &  \\
    \hline
    90/150 $\si{\GHz}$ & 1.42 & 427 & 2.28 & 270 & 0.8 \\
    \hline
    220/280 $\si{\GHz}$ & 1.42 & 210 & 2.49 & 123 & 0.2 \\
    \hline
    \end{tabular}
    \caption{Two-layer AR coating design parameters, index of refraction and thickness, and expected reflectance, averaged over the entire passband.}
    \label{table:parameters}
\end{table}

We demonstrate that two coating layers with index of refraction and thickness of each layer optimized for maximum transmission meets the performance requirements of Simons Array. Using the characteristic matrix formalism to calculate the transfer matrix at every boundary~\cite{hou}, we obtain the analytical solution for wave propagation through an arbitrary stack of dielectrics. The optimized transmittance spectra of two-layer quarter-wavelength AR coatings on alumina for the 90/150 and 220/280 $\si{\GHz}$ dual bands of Simons Array are shown in Figure~\ref{fig:opt_theory}. The parameters and band-averaged reflectance of the calculated ideal stack of dielectrics for these bands are shown in Table~\ref{table:parameters}. The 90/150 $\si{\GHz}$ spectrum shows reflection suppressed to below 5\% over 82\% fractional bandwidth, for 0.7, 0.8, and 0.8\% band-averaged reflectance at 90 $\si{\GHz}$, 150 $\si{\GHz}$, and total passband, respectively, and the 220/280 $\si{\GHz}$ spectrum shows reflection suppressed to below 5\% over 70\% fractional bandwidth, for 0.2, 0.1, and 0.2\% band-averaged reflectance at 220 $\si{\GHz}$, 280 $\si{\GHz}$, and total passband, respectively.

\begin{figure}[htbp]
\centering
\includegraphics[width=0.99\linewidth]{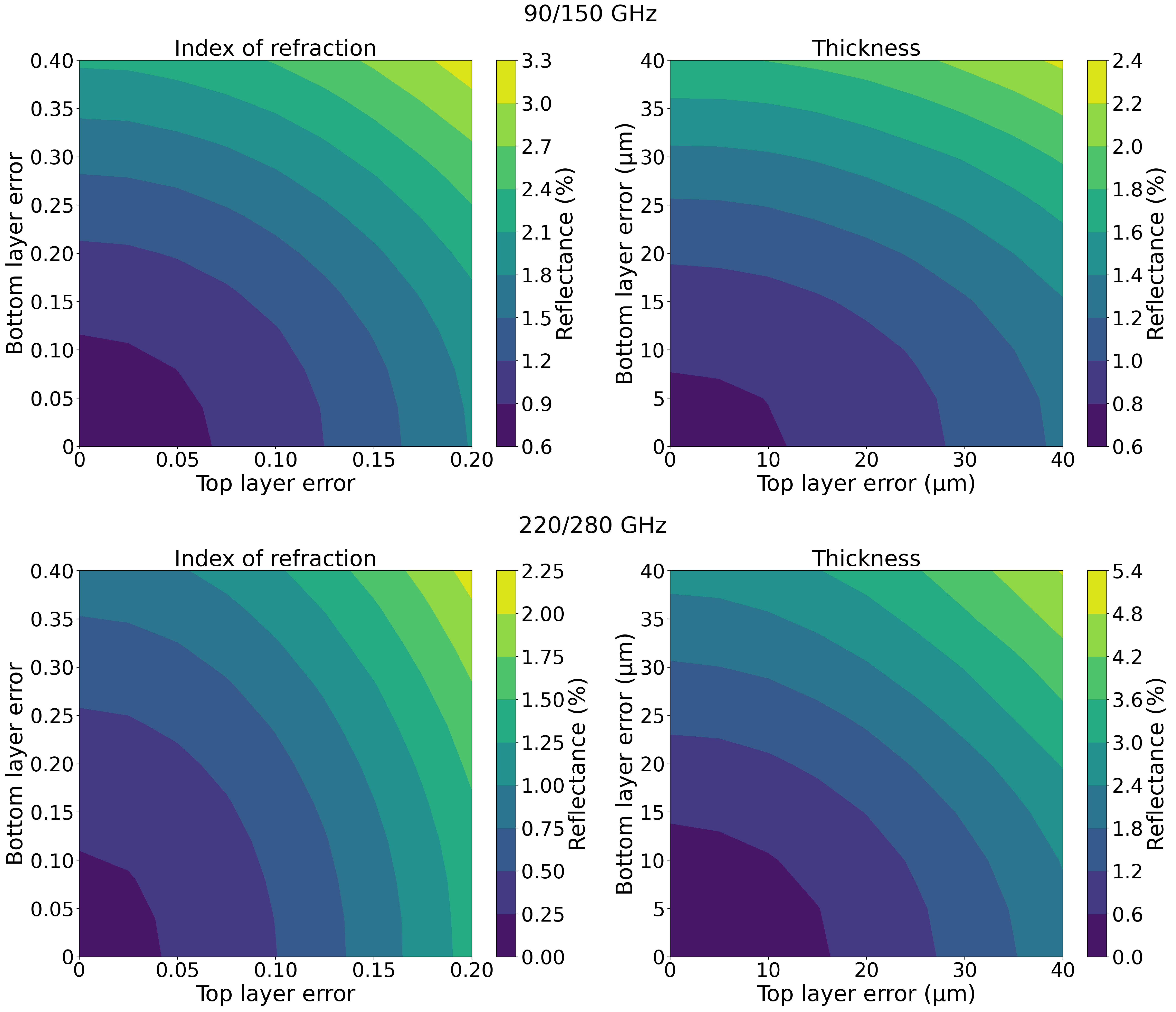}
\caption{Monte Carlo study on the effect of error in coating index of refraction and thickness on the efficiency of one optical element, AR coated for the 90/150 (top) and 220/280 $\si{\GHz}$ bands (bottom). Errors in index of refraction and thickness are simulated by averaging 500 trials of a normal distribution around optimal value with fixed standard deviation. The axes represent this standard deviation around the optimal values for a set of trials.}
\label{fig:tol}
\end{figure}

Errors in coating index of refraction and thickness lead to additional reflection in the passband. Shown in Figure~\ref{fig:tol}, we calculate the increase in total band-averaged reflectance from these errors in top and bottom layers of the two-layer AR coating by performing a Monte Carlo study. Reflectance is averaged across 500 trials with fixed standard deviation for index of refraction or thickness which are normally distributed around optimal values. This study shows an asymmetry in the effect of errors in top and bottom layers, most prevalent in index of refraction error. Furthermore, efficiency of the 220/280 GHz passband degrades more rapidly with coating errors than that of 90/150 GHz. We reference this study in Sections~\ref{sec:design} and~\ref{sec:fab} to evaluate additional loss due to uncertainty in index of refraction and thickness, respectively, from the fabrication method. 

\section{AR Coating Design}
\label{sec:design}

\begin{figure}[htbp]
\centering
\includegraphics[width=0.8\linewidth]{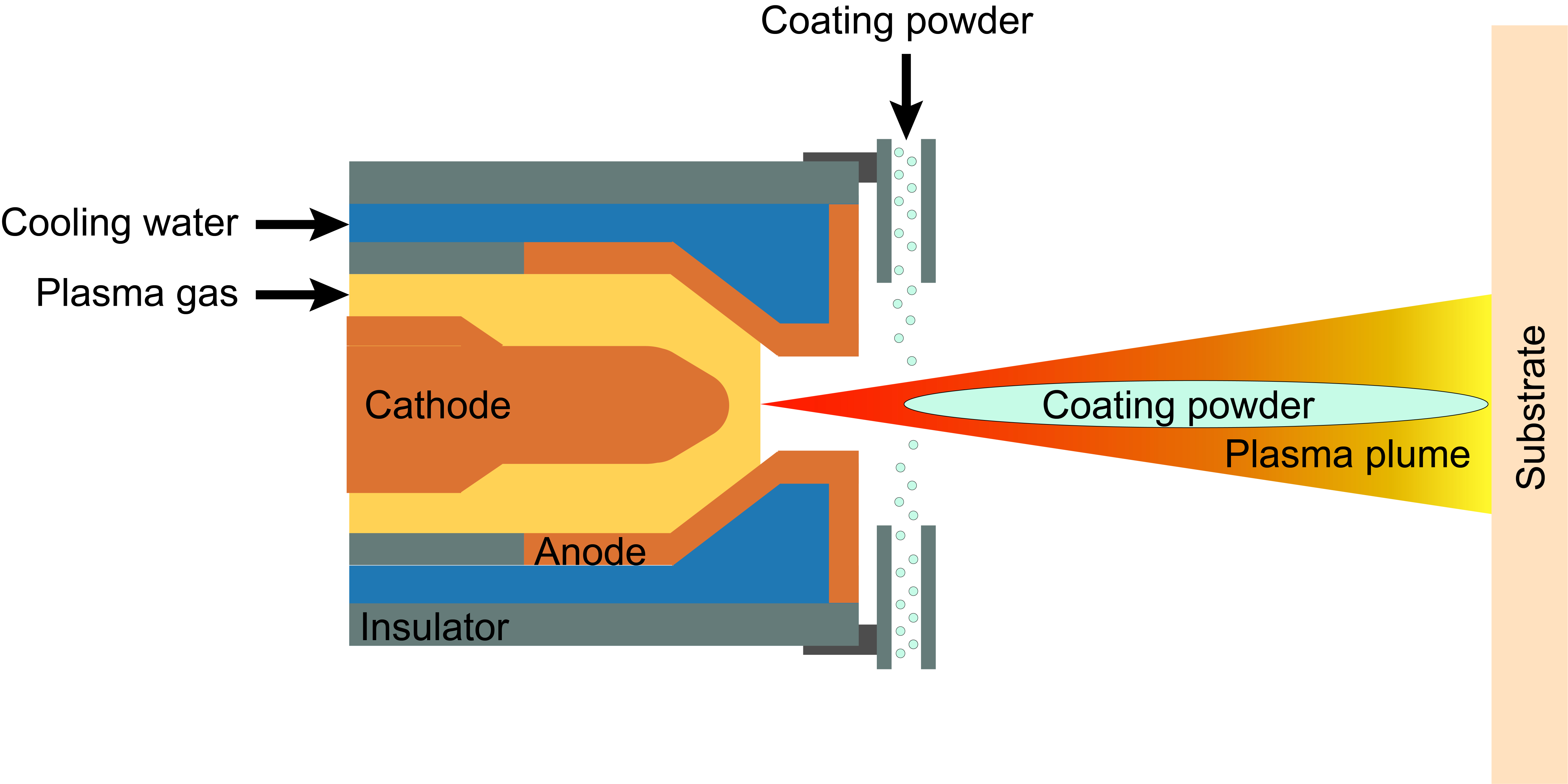}
\caption{Schematic diagram of a plasma spray gun. A powerful electrical arc ionizes gas to generate a plasma jet, which melts and propels the coating powder towards the substrate.}
\label{fig:spray_diagram}
\end{figure}

Plasma spray is a technique which injects coating feedstock material, such as alumina powder, into a direct current plasma jet with temperature of order $10^{5}$ $\si{\kelvin}$, heating it to a molten state and propelling it towards the substrate with high temperature and momentum. The direct current is applied to a chamber with a cathode-anode pair to generate a powerful electrical arc of $\sim 10^4$ $\si{\watt}$, as shown in Figure~\ref{fig:spray_diagram}. A gas flows continuously to this chamber and is ionized by the electrical arc, generating a plasma plume. The molten stream cools upon contact with the surface, mechanically cross-linking with itself and the substrate to form a strongly adhered coating without any glue or adhesive chemical. To promote adhesion, the substrate is grit-blasted to obtain surface roughness of approximately 5 $\si{\um}$ rms. 

We utilize an argon plasma jet with alumina- and silica-based feedstock powder. Three distinct powders are used to form the set of feedstock material: Oerlikon Metco 6051, HAI HA008, and Tolsa Fillite 160W which are solid alumina, hollow alumina microsphere, and hollow silica microsphere powders, respectively. To develop a technology with tunable index of refraction, we vary the following parameters: plasma power (arc voltage $\times$ arc current), powder feed rate, stand-off distance, plasma gas type, and powder matrix composition~\cite{thirumalaikumarasamy}. 

\begin{figure}[htbp]
\centering
\includegraphics[width=0.7\linewidth]{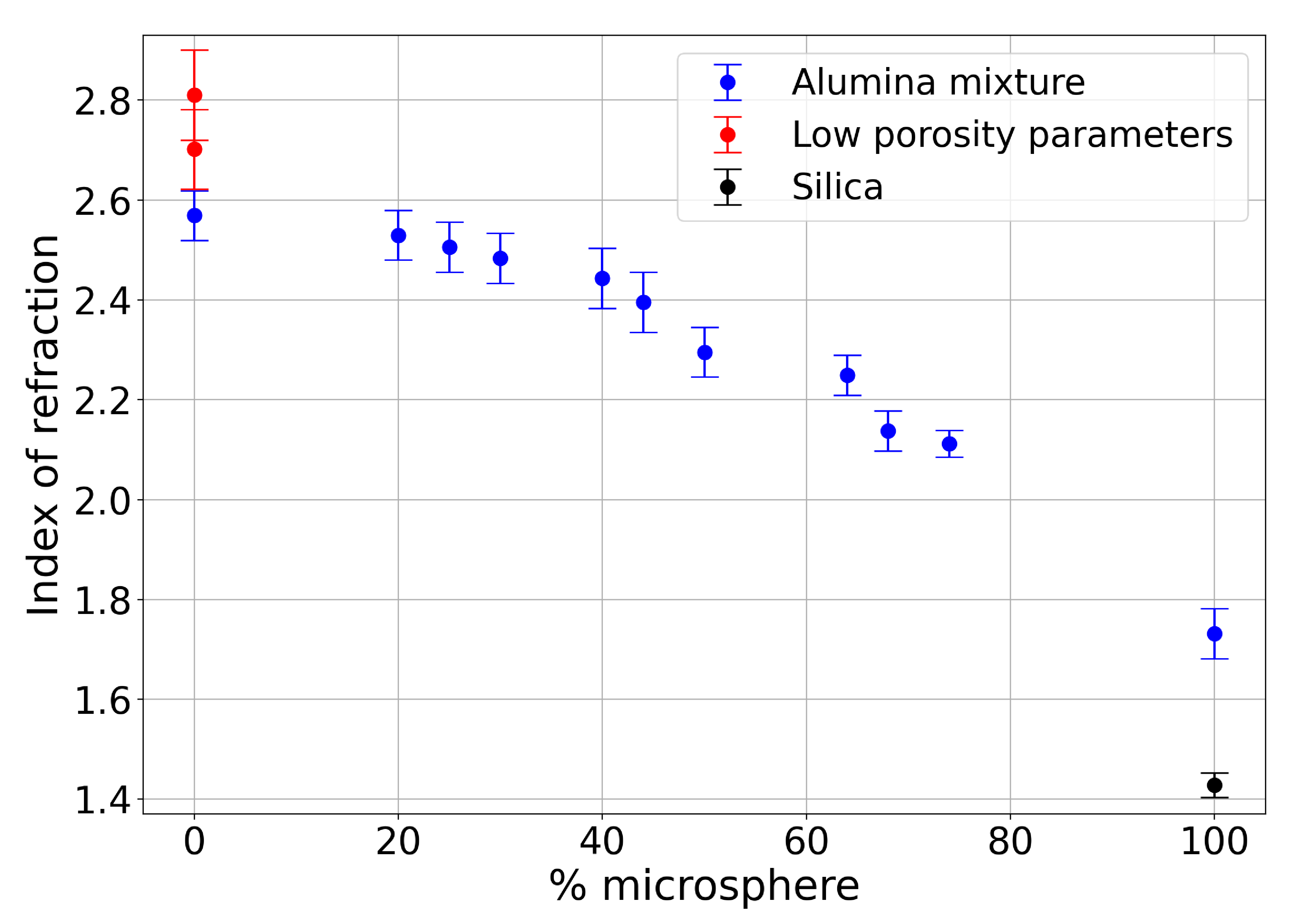}
\caption{Range of achievable index of refraction, produced by varying the powder matrix and plasma spray parameters.}
\label{fig:DC_range}
\end{figure}

\begin{figure}[htbp]
\centering
\includegraphics[width=0.7\linewidth]{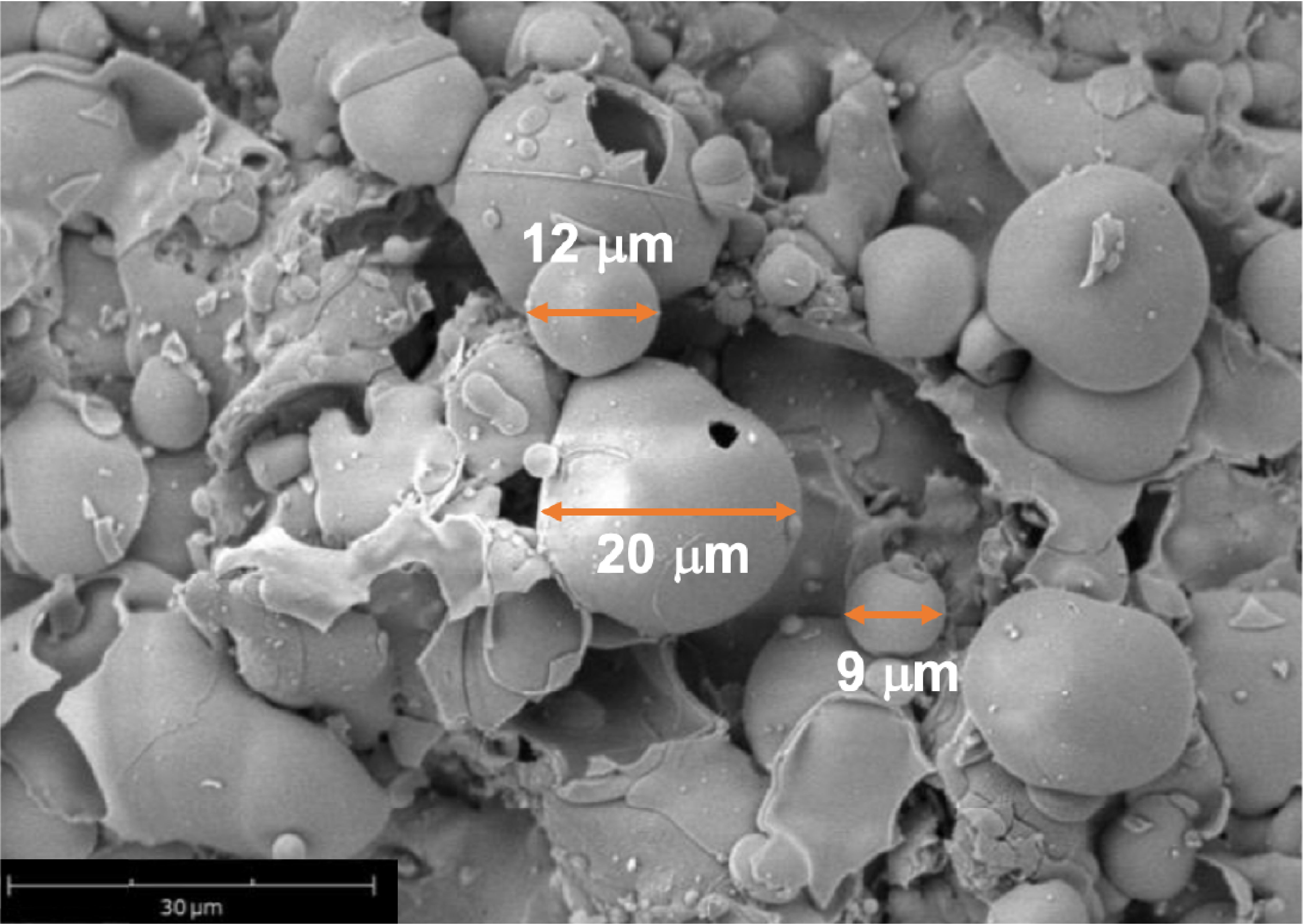}
\caption{Cross-section image of a plasma sprayed coating, taken using a scanning electron microscope at 3700$\times$ magnification, 5 $\si{\kV}$ accelerating voltage, and 5.768 $\si{\mm}$ working distance. The sample is prepared by breaking a fragment from a sprayed coating. The annotated particles show the thin-walled HA008 microspheres survive the plasma spray process and produces a coating with porous bulk structure.}
\label{fig:sem}
\end{figure}

Figure~\ref{fig:DC_range} presents the achievable range of index of refraction of this technology, using the measurement process described in Section~\ref{sec:setup}, and is referenced to determine the powder matrix and spray parameters for the desired index of refraction. By varying the composition of Metco 6051 and HA008, we span the range of indices between 1.73 $\leq$ n $\leq$ 2.59 (blue points in Figure~\ref{fig:DC_range}). Higher concentration of hollow microspheres leads to proportionally higher porosity and thus lower index, and vice-versa. Further adjustments are made to push the lower and upper bounds of achievable indices. To produce n = 1.42 for the top layer of an optimal two-layer AR coating, we spray a powder matrix composed solely of Fillite 160W due to the relatively low index of silica, and adjust the powder feed rate and stand-off distance for colder and less energetic powder (black point in Figure~\ref{fig:DC_range}). To produce higher values of index of refraction up to n = 2.81, we spray a powder matrix composed solely of Metco 6051, and increase the plasma power and plasma gas flow rate for hotter and more energetic powder (red points in Figure~\ref{fig:DC_range}). Furthermore, hydrogen gas is used as a secondary gas, as the dissociation of its diatomic molecule contributes additional energy to the powder. The cross-sectional image of these coatings taken using a scanning electron microscope, shown in Figure~\ref{fig:sem}, shows the micropores inherent to the coatings as a result of utilizing hollow microsphere powders. Microsphere powders utilized in this paper are chosen with significantly smaller grain sizes than the wavelengths of interest to minimize scattering loss.

Due to the wide range of available indices, a variety of AR coating designs can be sprayed from single- to multi-layer designs. For the 90/150 $\si{\GHz}$ bands of Simons Array, we spray a powder matrix consisting of 52\% Metco 6051 and 48\% HA008 to construct the bottom layer. Then we spray a powder matrix consisting only of Fillite 160W to construct the top layer for a complete plasma sprayed AR coating. For the 220/280 $\si{\GHz}$ bands of Simons Array, we spray a powder formulation of 32\% Metco 6051 and 68\% HA008 for the bottom layer while the top layer parameters are identical to that of the 90/150 $\si{\GHz}$ design.

\par

For both passbands, the uncertainty of the top layer index is $\pm$0.02 and that of the bottom layer indices is $\pm$0.05. This leads to 0.2\% increase in reflectance of the 90/150 and 220/280 GHz AR coatings, which are well within the tolerance of $\leq$1\% additional loss for a single AR coated optical element, discussed in Section~\ref{sec:req}.

\section{Fabrication Process}
\label{sec:fab}

\begin{figure}[htbp]
\centering
\includegraphics[width=0.5\linewidth]{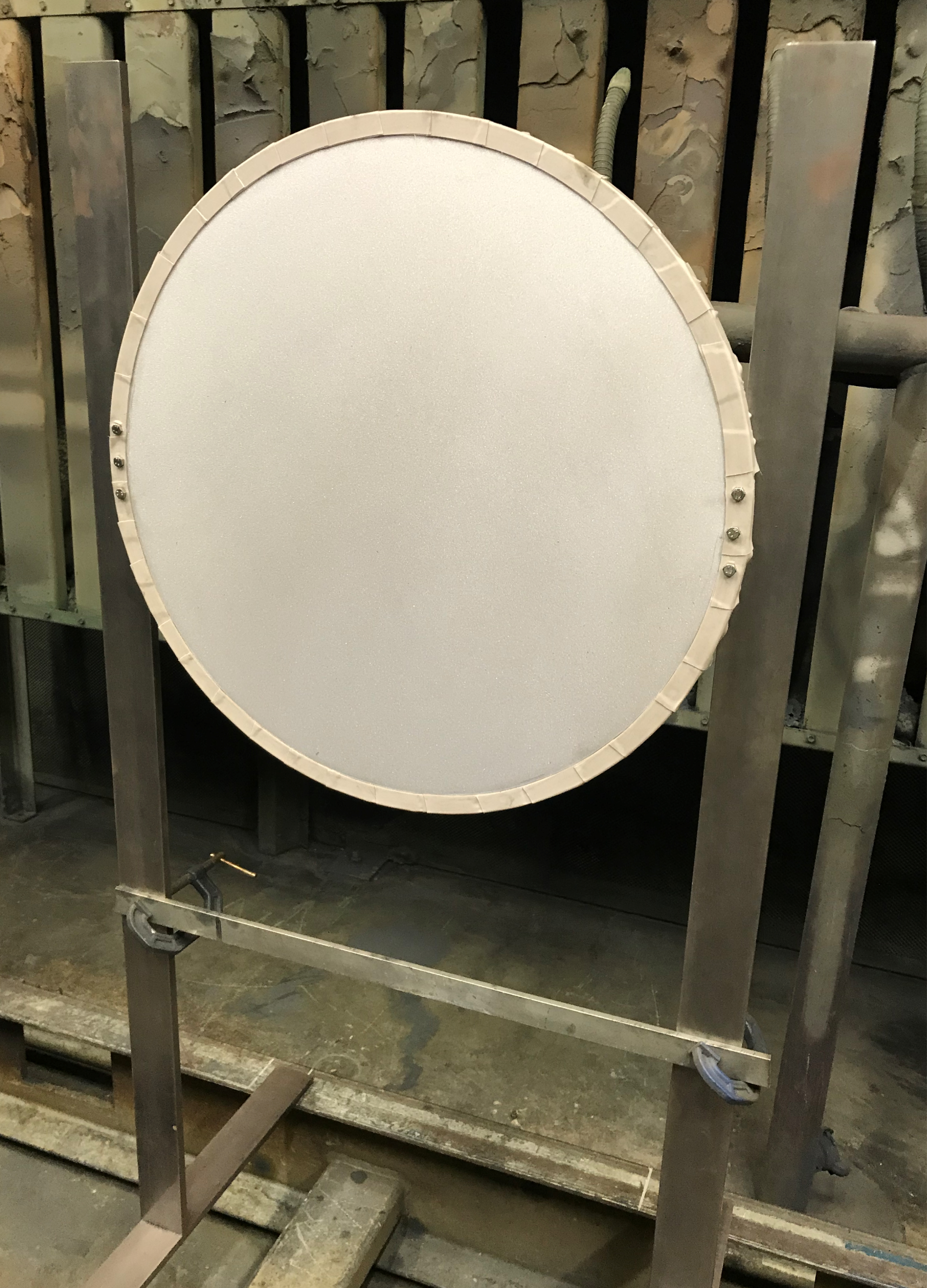}
\caption{POLARBEAR-2b lens spray coated on a fabrication fixture.}
\label{fig:lens_fixture}
\end{figure}

Using a plasma spray gun mounted on a robot manipulator, an object of complex geometry can be coated with high precision and accuracy. The robot manipulator is programmed to trace the surface of the aspheric lens with constant angle and stand-off distance due to their strong influence on coating deposition rate and index~\cite{tillmann, ilavsky, leigh}. The target substrate is securely held using a fixture during the robot programming and spray process. An example of a POLARBEAR-2b lens on a fixture is shown in Figure~\ref{fig:lens_fixture} with its optical axis pointing perpendicular to gravity. The maximum diameter that can be coated is constrained by the limits of the robot travel distance, which was 2000 $\si{\mm}$ for the system used in this paper.

The robot is programmed to move in a raster scan pattern over the substrate surface throughout each spray deposition pass until the desired thickness is achieved. The index of refraction has a strong dependence on raster step size and traverse speed, and thus they are fixed at 3 $\si{\mm}$ and 750 $\si{\mm}/\si{s}$ throughout the production process. For a convex surface, each raster scan line follows a circular arc of constant traverse speed, defined by the lens height at three points: two at the edge of the optical surface and one at the center corresponding to the highest point. This allows the plasma gun to spray along the raster scan line at constant stand-off distance. Furthermore, the manipulator system adjusts the angle of the gun to be normal to the arc of the raster scan to maintain a spray angle within $\pm \ang{10}$ of the normal vector to the substrate surface. Turnaround points are placed far from the edge points in order to maintain uniform traverse speed and thus deposition rate over the substrate surface. 

\par

\begin{figure}[htbp]
\centering
\includegraphics[width=0.99\linewidth]{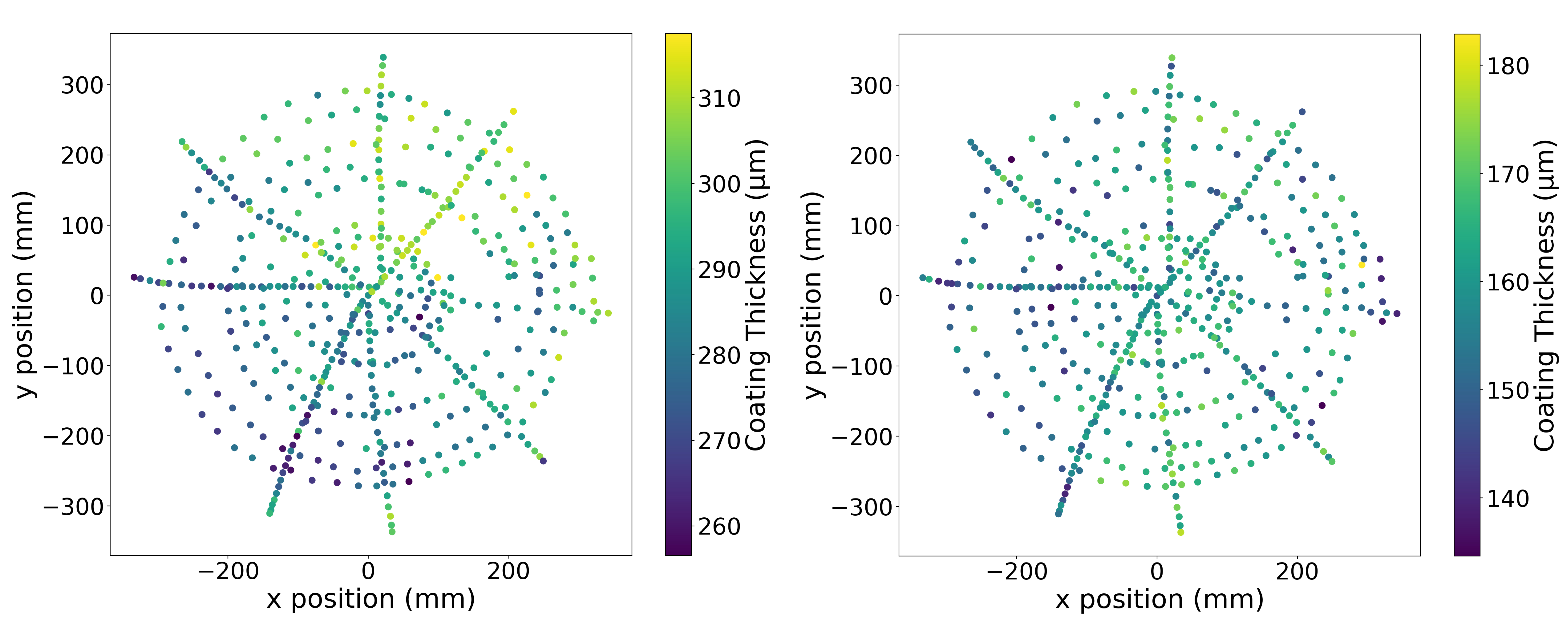}
\caption{CMM measurements of Metco 6051/HA008 coating thickness (left) and Metco 6051 coating thickness (right) over a 700 $\si{\mm}$ diameter, convex alumina surface shows $\pm 7$ $\si{\um}$ and $\pm 10$ $\si{\um}$ 1-$\sigma$ uniformity, respectively.}
\label{fig:cmm_metrology}
\end{figure}

Figure~\ref{fig:cmm_metrology} shows coordinate measuring machine (CMM) measurements of achieved thickness value and uniformity of a two-layer AR coating on the 700 $\si{\mm}$ diameter, convex surface of an SPT-3G alumina lens~\cite{sobrin}. The thickness of both coating layers are within 10 $\si{\um}$ of the target values. The coating layer of Metco 6051 achieves tighter thickness uniformity of $\pm$7 $\si{\um}$ while that of the Metco 6051/HA008 matrix achieves uniformity of $\pm$10 $\si{\um}$. Precision of this measurement process is $\pm$3 $\si{\um}$, determined by measuring the same, uncoated aspheric lens surface twice in a similar fashion to determining the uniformity of coating thickness. In between two measurements, the bare lens is removed and then placed again on the CMM for re-alignment. Thickness uniformity of $\pm$10 $\si{\um}$ for both layers contribute 0.3 and 0.4\% increase in reflectance of the 90/150 and 220/280 GHz AR coatings, respectively, which are well within the tolerance of $\leq$1\% additional loss for a single AR coated optical element.

The powder matrix preparation and spray fabrication process require a dry environment due to the effect of water vapor on the powder flow of the feed system~\cite{stanford} and measured index of the coating~\cite{niittymaki}. To facilitate stable powder injection into the plasma, the powder matrix is baked for 2 hours at $\ang{130}$C and kept from exposure to the humid atmosphere. A dehumifier is in continuous operation during the spray process to reduce relative humidity below 30$\%$.  

\par

Following this fabrication procedure, it typically takes 1 day to coat both sides of an optical element with one layer of coating. While it does not take more than 1--2 hours for the spray deposition, the overall production duration is lengthened by periodic measurements of coating thickness and setup time of equipment, powders, and optical element.

\par

All coupons and lenses shown in this paper were sprayed by the commercial vendor, Curtiss-Wright Surface Technologies. Most atmospheric plasma spray equipment systems available in commercial facilities can adopt this technology due to the standard nature of its equipment set and the spray parameters. While a set of commonly used 3M equipment is used for this paper, another set can be used with adjustments to the spray parameter set to produce comparable results.

\section{Measurement Setup}
\label{sec:setup}

\begin{figure}[htbp]
\centering
\includegraphics[width=0.99\linewidth]{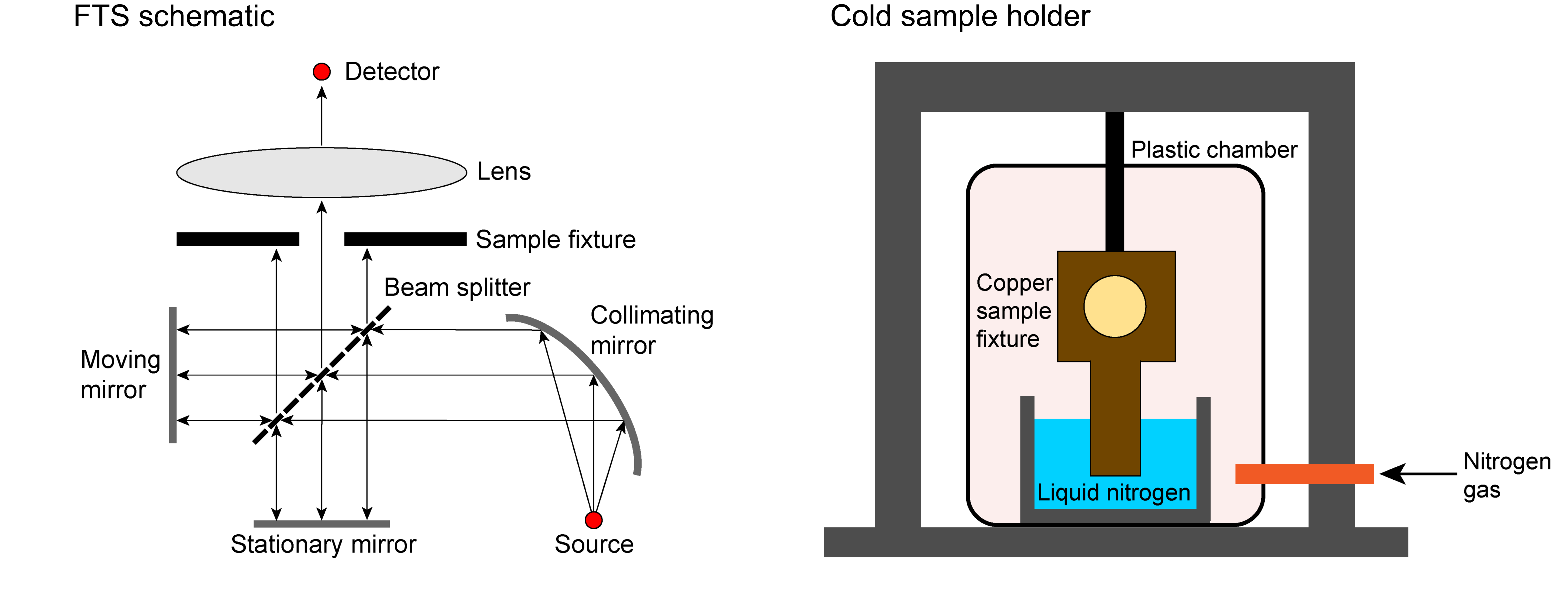}
\caption{Schematic diagrams of the Michelson FTS (left), used to measure the transmittance spectrum of 51 $\si{\mm}$ diameter coupons, and the cold sample holder (right), used to cool sample coupons to 100 $\si{\kelvin}$ for cryogenic measurements.}
\label{fig:fts}
\end{figure}

\begin{figure}[htbp]
\centering
\includegraphics[width=0.58\linewidth]{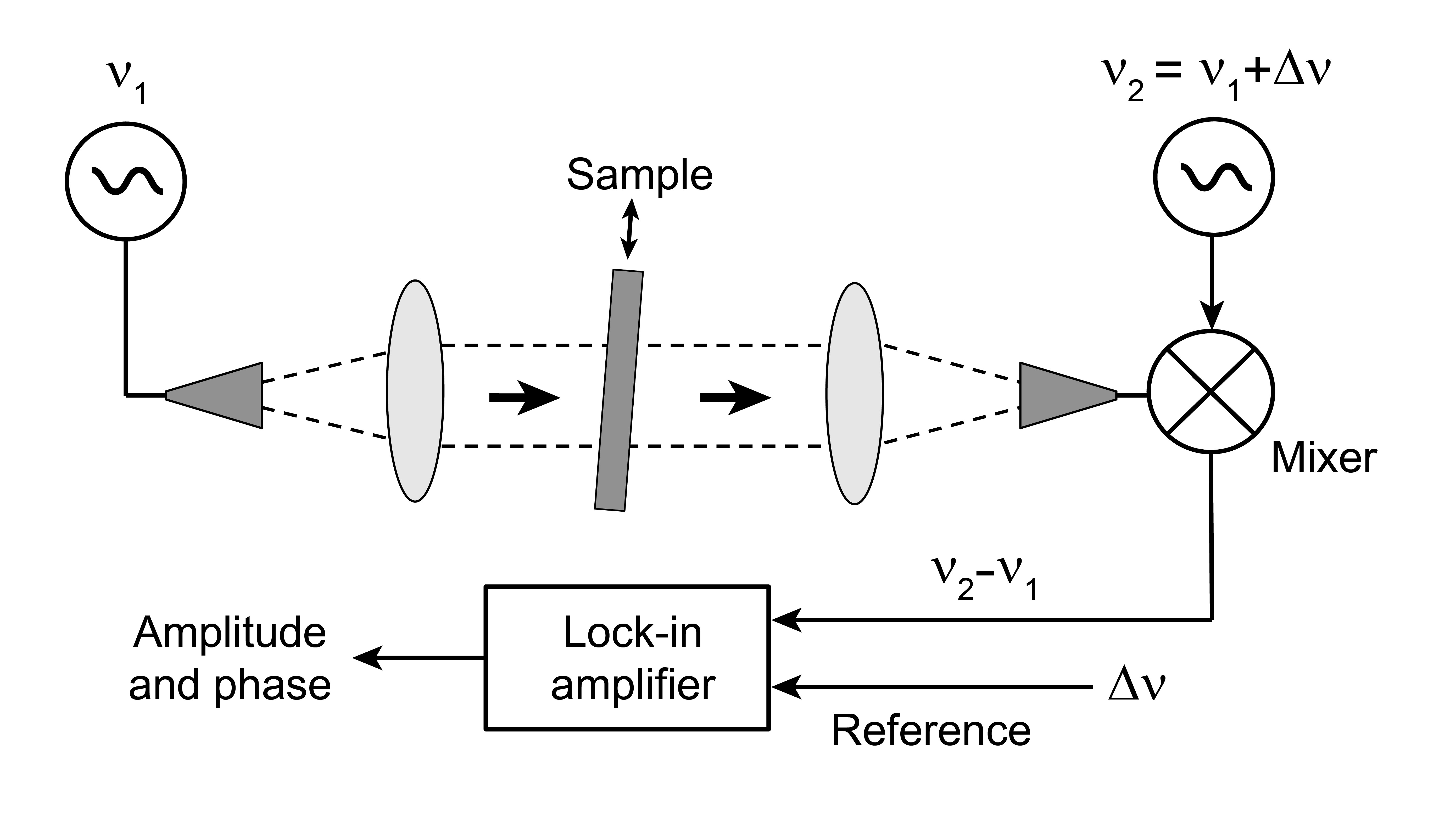}
\caption{Schematic diagram of the heterodyne spectrometer, used to measure the transmittance spectrum of 51 $\si{\mm}$ diameter coupons.}
\label{fig:het}
\end{figure}

We measure the optical properties of samples using the Michelson Fourier Transform Spectrometer (FTS) and and the heterodyne spectrometer, whose schematic diagrams are shown in Figures~\ref{fig:fts} and~\ref{fig:het}. The FTS measures the spectral response of a dielectric sample to determine its optical properties (see~\cite{rosen} for detailed description). The detector is a filter-less, antenna-coupled transition-edge sensor bolometer which is read out using a superconducting quantum interference device~\cite{obrient, westbrook}. This system provides sensitivity between 80--300 $\si{\GHz}$  with a typical resolution of 1 $\si{\GHz}$. The transmittance spectrum of a sample is measured by dividing the spectral response of the detector with the sample ("closed") in the optical path by the spectral response without the sample ("open"). In order to improve the signal-to-noise, we average over multiple pairs of open and closed measurements. To minimize the impact of gain drift on the normalization of an averaged transmittance spectrum, the time-ordering of an open and closed measurement is alternated from pair to pair. 

Figure~\ref{fig:fts} shows the sample fixture which cools the sample to 100 $\si{\kelvin}$ for cryogenic measurements. This cooling system consists of a copper sample fixture which is coated with an Eccosorb absorbing screen around the 51 $\si{\mm}$ aperture and partially immersed in liquid nitrogen below the path of the beam. To prevent condensation on the sample which absorbs millimeter-wave light, the copper aperture and liquid nitrogen can are placed inside a transparent chamber where dry nitrogen continuously flows. In cryogenic measurement configuration, the sample fixture is placed immediately after the collimator mirror due to space constraints.

The heterodyne spectrometer measures the change in the amplitude and phase of a monochromatic signal when the dielectric sample is inserted into the beam. The sample is placed at a slight angle to reduce the effects of standing waves. The probe signal is down-converted to a lower frequency in a mixer, then synchronously detected with a lock-in amplifier (see~\cite{schultz} for more detailed description). The probe signal from a tunable oscillator~\cite{plambeck} is stepped over a range of frequencies, typically 70--114 $\si{\GHz}$ in 0.3 $\si{\GHz}$ steps. Though the heterodyne spectrometer covers a narrower frequency range than the FTS, the inclusion of phase information helps to constrain fits of optical properties and is used to corroborate FTS measurements.

\begin{figure}[htbp]
\centering
\includegraphics[width=0.99\linewidth]{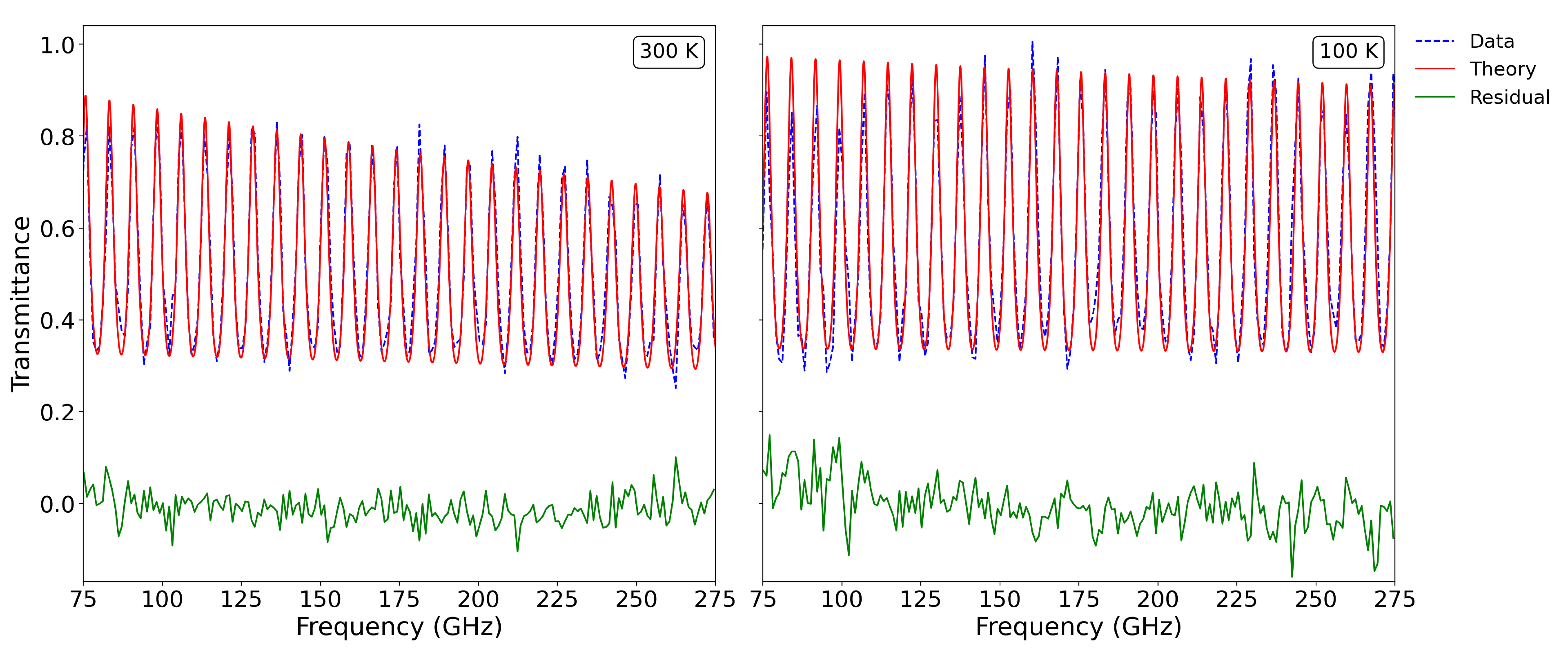}
\caption{Transmittance spectrum measurements of an alumina coupon at 300 (left) and 100 $\si{\kelvin}$ (right). This figure shows FTS data (blue), model (red), and residual (green).}
\label{fig:alumina_fts}
\end{figure}

For Simons Array, a rigorous process of validation was implemented in which each optic's dielectric properties and shape were measured and fed into Zemax optical design software containing the Simons Array optics chain to ensure sufficient coupling between the lenses and telescope. A 51 $\si{\mm}$ diameter, 6.350 $\si{\mm}$ thick witness coupon was cut from each lens blank of 99.9\% purity alumina, prior to cutting to aspheric shape, and measured with the FTS. Figure~\ref{fig:alumina_fts} displays an FTS transmittance spectrum measurement and fit of one such witness coupon at 300 and 100 $\si{\kelvin}$. The thickness of the coupon is measured at both temperatures using a micrometer and is known to $\pm$0.002 $\si{\mm}$. Given the degeneracy between thickness and index of refraction in the amplitude and rate of Fabry-Perot fringes, uncertainty in coupon thickness limits the precision in the fitted value of index of refraction. The model of best fit shows this alumina is characterized by n = 3.110 $\pm$ 0.001 and tan $\delta$ = (6.3 $\pm$ 0.3) $\times$ 10$^{-4}$ at 300 $\si{\kelvin}$ and n = 3.108 $\pm$ 0.001 and tan $\delta$ = (9 $\pm$ 4) $\times$ 10$^{-5}$ at 100 $\si{\kelvin}$.

\section{Performance}
\label{sec:performance}

\begin{figure*}[htbp]
\centering
    \includegraphics[width=0.95\textwidth]{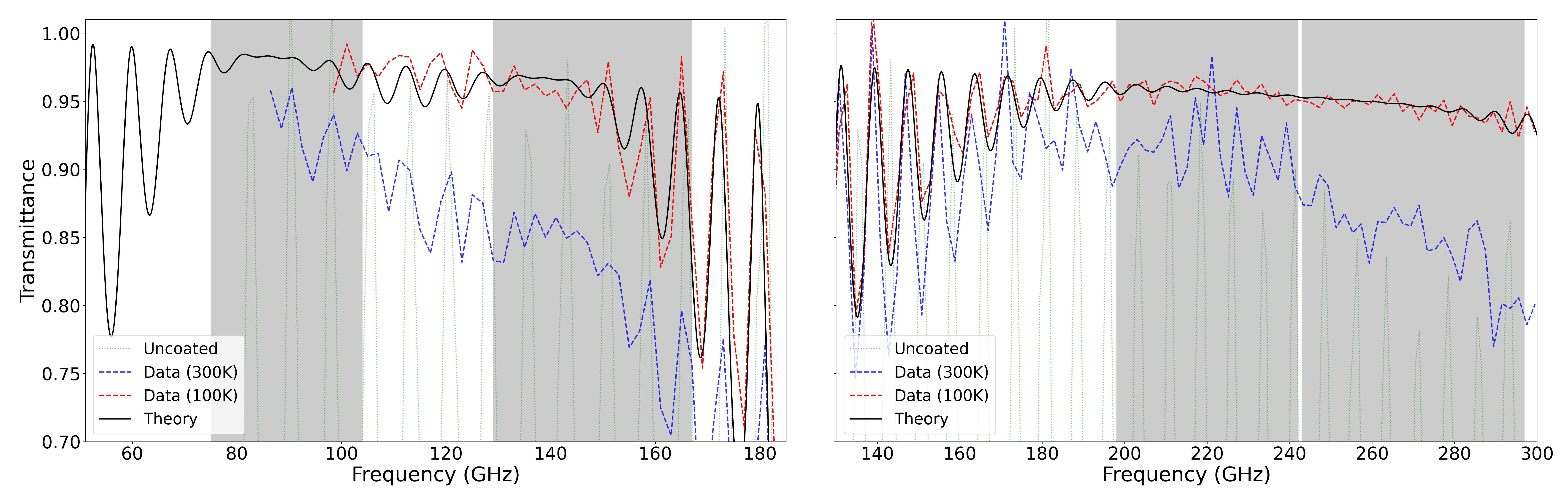}
    \caption{Transmittance spectrum measurements of alumina coupons coated for optimal performance at 90/150 $\si{\GHz}$ bands (left) and 220/280 $\si{\GHz}$ bands (right). This figure shows FTS data of alumina at 300 $\si{\kelvin}$ (green), FTS data of AR coated alumina at 300 (blue) and 100 $\si{\kelvin}$ (red), and model to the FTS data of AR coated alumina at 100 $\si{\kelvin}$ (black).}
    \label{fig:tx_fts}
\end{figure*}

We sprayed alumina coupons with AR coating designs described in Section~\ref{sec:design} for the 90/150 and 220/280 $\si{\GHz}$ passbands. The transmittance of these samples were measured using the FTS as shown in Figure~\ref{fig:tx_fts} at 300 and 100 $\si{\kelvin}$. Reflectance is derived from these measurements by setting the loss tangent of every dielectric layer in the best fit model to zero, such that only the oscillatory component of the wave propagation equation remains and reflection is the sole source of loss. Both band-averaged reflectance and fractional bandwidth are calculated from this method. The 90/150 $\si{\GHz}$ sample measurement shows reflection suppressed to below 5\% over 82\% fractional bandwidth, with 0.4, 2.4, and 1.6\% band-averaged reflectance for 90 $\si{\GHz}$, 150 $\si{\GHz}$, and total passband, respectively, and the 220/280 $\si{\GHz}$ sample measurement shows reflection suppressed to below 5\% over 69\% fractional bandwidth, with 0.1, 0.3, and 0.2\% band-averaged reflectance for 220 $\si{\GHz}$, 280 $\si{\GHz}$, and total passband, respectively. Cooling the samples reduces the band-averaged absorption loss from 13\% and 12\% to $\sim$1\% for the 90/150 $\si{\GHz}$ and 220/280 $\si{\GHz}$ AR coatings, respectively. Since re-imaging lenses for CMB experiments typically operate at 4 K, we expect further reduction in absorption loss at operating temperatures.

\par

Plasma sprayed AR coating has demonstrated robustness to cryogenic delamination for 6.35 $\si{\mm}$ radius alumina hemisphere lenslets, 51 $\si{\mm}$ diameter alumina coupons~\cite{jeong}, 500 $\si{\mm}$ diameter alumina plate, 500 $\si{\mm}$ diameter alumina refractive elements within the POLARBEAR-2b receiver~\cite{lowry}, and 700 $\si{\mm}$ diameter alumina refractive elements within the SPT-3G receiver~\cite{sobrin}. The smaller lenslets and coupons have each been thermal cycled by liquid nitrogen submersion 10 times. The 500 $\si{\mm}$ plate has been thermal cycled by liquid nitrogen submersion 3 times. The refractive elements of POLARBEAR-2b and SPT-3G were each thermal cycled by pulse tube cooling to 4 $\si{\kelvin}$ a single time.

\section{Conclusion}
\label{sec:conclusion}

We have developed an AR coating technology using plasma spray for alumina and silicon refractive elements over a broad bandwidth. Coatings of index of refraction between 2.0 and 7.9 are sprayed with order 10 $\si{\um}$ accuracy and precision over a large diameter with matching CTE to alumina substrates. We demonstrated two-layer AR coatings which suppress reflection from alumina to below 5\% over 82\% and 69\% fractional bandwidths for 90/150 and 220/280 $\si{\GHz}$ passband designs, respectively. 

\par

This broadband AR coating is applicable for alumina and silicon refractive optics of millimeter-wave astronomy, and can be expanded to wider bandwidths with additional layers. Optimal AR coating on alumina with more than two layers requires a top layer index below what is currently achievable by this technology. However, a multi-layer design utilizing the range of indices shown in Figure~\ref{fig:DC_range} can yield a sufficiently low reflectance. For example, a four-layer AR coating designed to simultaneously cover the 90/150/220/280 GHz bands of Simons Array will achieve 1.9\% total band-averaged reflectance, with indices 1.42, 1.88, 2.42, and 2.81 from top to bottom of the coating stack. 

\par

Implementations of this technology to frequency bands beyond 30--500 GHz require further adjustments to the spray parameters and fabrication procedure. Low frequency coverage requires coating thicknesses of order 1 $\si{\mm}$. Delamination can occur during spray deposition of such thickness as a large temperature gradient develops along the thickness of the coating and places significant mechanical stress. High frequency coverage requires smaller thicknesses, for which additional loss from thickness non-uniformity will increase towards ~1\%. Furthermore, scattering rate scales with frequency and will require microspheres of smaller grain size.  

\section*{Acknowledgement}

The work presented here and Simons Array are supported by the Simons Foundation, the National Science
Foundation (AST-1440338), the Moore Foundation, and the Templeton Foundation.

% Bibliography
\bibliography{references}

\end{document}